\documentclass[12pt,fleqn]{article}
\usepackage[margin=3.5cm]{geometry}
\usepackage{lipsum} 
\usepackage{amsmath}
\usepackage{slashed}
\usepackage{url}
\usepackage{bm}
\usepackage{graphicx}
\usepackage[utf8]{inputenc}
\usepackage[english]{babel}
\usepackage{amsmath}
\usepackage{amsfonts}
\usepackage{amssymb}
\usepackage{float}
\usepackage{hyperref}
\usepackage{extarrows}
\usepackage{fullpage}
\begin{document}
\linespread{1.5}
\renewcommand{\baselinestretch}{1.2}
  \fontsize{13}{16}\selectfont

\title{Yang--Mills Theory of Gravity}
\author         {Malik AL Matwi}

\date{
Department of Mathematical Science, Ritsumeikan University, 4-2-28, Kusatsu-Shi, Shiga, 525-0034, Japan\\ malik.matwi@gmail.com\\ 
This version includes some corrections to the \\
published paper: physics 2019, 1(3), 339-359}

\maketitle
\tableofcontents

\section*{Abstract}
The canonical formulation of general relativity is based on decomposition space--time manifold $M$ into $ R\times \Sigma$, this decomposition has to preserve the invariance of general relativity, invariance under general coordinates, and local Lorentz transformations. These symmetries associate with conserved currents that are coupled to gravity. In this paper, we try  to solve the equations of motion of general relativity in self-dual formalism using only the spin currents(Lorentz currents), in static case, and without needing using the Einstein's equation, that makes the general relativity similar to Yang-Mills theory of gauge fields. We give an example, matter located at a point, so we have spherical symmetric system. Then we add Yang--Mills Lagrangian to general relativity Lagrangian. Finally we use the decomposition of the space--time manifold $M=R\times \Sigma$ to find that $\Sigma^{0a}_i$ is a conjugate momentum of $A^{i}_a$ and $\Sigma^{ab}_i F_{ab}^i$ is energy density.
\\

Keywords: Lorentz invariance; spin currents; complex connection $A^i$.

\section*{Introduction}
Gravity can be formulated based on gauge theory by gauging the Lorentz group $SO(3,1)$ \cite{Merced}. For this purpose, we need to fix some base space and consider that the Lorentz group $SO(3,1)$ acts locally on Lorentz frames which are regarded as a frame bundle over a fixed base space $M$. We can consider this base space as an arbitrary space--time manifold $M$ with coordinates $(x^\mu)$, and consider the local Lorentz frame as an element in the tangent frame bundle over $M$. By that we have two symmetries; invariance under continuous transformations of local Lorentz frame, $SO(3,1)$ group, and invariance under diffeomorphism of the space--time $M$, which is originally considered as a base space \cite{James}.
\\

The local Lorentz invariance gives a conserved current, call it spin current, or Lorentz current. In flat coordinates $(x^I)$, the spin current for arbitrary field $\varphi$ is (\cite{Mark}, section 22)
\[
M^{IJK}=x^J T^{IK} - x^K T^{IJ},
\]
the conservation law is $\partial_I M^{IJK}=0$, with conservation of energy-momentum tensor $\partial_I T^{IJ}=0$. When the field carries spinor indices, like $\varphi^\alpha$, this adds a term like
\[
M^{IJK}=x^J T^{IK} - x^K T^{IJ} + \pi_\alpha^I (S^{JK})^\alpha{_\beta} \varphi^\beta.
\]
In arbitrary coordinates $(x^\mu)$, we write $(x^I=e^I_\mu x^\mu)$, where $x^I$ becomes tangent vector, also $T^{\mu\nu}=e_I^\mu e_J^\nu T^{IJ}$. One can write
\[
M^{\mu IJ}=e^I_\nu x^\nu T^{\mu J} - e^J_\nu x^\nu T^{\mu I} .
\]
Therefore $D_\mu M^{\mu IJ}=0$ when $D_\mu e^I_\nu=0$ and $D_\mu T^{\mu I}=0$. The current $M^{\mu IJ}$ couples to the spin connection $\omega^{IJ}_\mu$ for local symmetry of Lorentz group $SO(3,1)$. The equation $D_\mu e^I_\nu=0$ defines affine connection $\nabla$ on $TM\times TM$ and spin connection $\omega$ on $TM\times \mathfrak{so}(3, 1)$. In this paper, we try to solve the equation of motion of the Plebanski Lagrangian 
\begin{equation}\label{eq:z9}
L= B^{\mu\nu }_i F^i_{\mu\nu }  +\phi_{ij} B^{\mu\nu i}B^j_{\mu\nu } +L_{matter},
\end{equation}
using only the spin connection, without needing using the affine connection $\nabla$(avoiding using it).

\section{Solutions using spin currents}
By regarding the local Lorentz symmetry as a gauge symmetry with spin connection $\omega \in \Omega^1(M, \mathfrak{so}(3, 1))$ (or $A\in \Omega^1(M, \mathfrak{so}(3, \mathbb{C}))$) as gauge fields, we recognize Yang--Mills theory in gravity. But not full gravity, since in the Yang--Mills theory, the variables are connections and conserved currents, while in the gravity the metric is also variable. The local Lorentz symmetry produces locally conserved currents, and those currents are coupled to spin connection $\omega^{IJ}$. This makes the local Lorentz symmetry a gauge symmetry with the Lorentz group as a gauge group. Also, these currents must be conserved and vanish in the vacuum. We try in this section to solve the equations of motion of general relativity in self-dual formalism using only the spin current(Lorentz current), in static case, and without needing solving the Einstein's equation. We give an example, matter located at a point, so we have spherical symmetric system.
\\

In self-dual formalism of general relativity, the independent variables are $B^{\mu\nu }_i$ and $A^i_\mu$, they take values in Lie algebra $\mathfrak{so}(3, \mathbb{C})$. The non-zero spin current $\delta L_{matter}/\delta A^i_\mu$ implies $D_\mu B^{\mu\nu i }\ne 0$(here $D_\mu=\nabla_\mu+A_\mu$). Therefore we write $B ^{\mu\nu i }= K{^i}_j \Sigma^{\mu\nu i }$, for some $3\times 3$ complex matrix $ K{^i}_j$ of scalar functions on $M$ in order to get $D_\mu \Sigma^{\mu\nu i }=0$ and $\Sigma^{\mu\nu i }=P^{iIJ} e_I^\mu e_J^\nu$, with gravitational field $e^I_\mu$ satisfying $D_\nu e^I_\mu=0$, here $D_\mu=\nabla_\mu +\omega_\mu$, where $\nabla$ is affine connection on $M$, we avoid using it. The gravitational field $e^I_\mu$ defines a metric $g_{\mu \nu }=e^J_\mu  e_{J\nu }$, it satisfies $\nabla _\rho g_{\mu \nu }=(D_\rho e^J_\mu)  e_{J\nu }+ e^J_\mu  (D_\rho e_{J\nu })=0$. The field $e_I^\mu$ is inversion of $e^I_\mu$, that is $e^I_\mu e_J^\mu=\delta^I_J$ and $e^I_\mu e_I^\nu=\delta^\nu_\mu$.
\\

We try to find the solutions of $K{^i}_j $ and $\Sigma^{\mu\nu i }$, so obtaining the spin connection $A^i_\mu$ from $D_\mu \Sigma^{\mu\nu i }=0$. By that we obtain the curvature of $A^i_\mu$ and the field $B ^{\mu\nu i }= K{^i}_j \Sigma^{\mu\nu i }$.
\\

The general relativity Lagrangian in self-dual formalism including matter is
\begin{equation}\label{eq:z9}
 L_{dual\text{ } GR}+L_{matter}= B^{\mu\nu }_i F^i_{\mu\nu }  +\phi_{ij} B^{\mu\nu i}B^j_{\mu\nu } +L_{matter},
\end{equation}
where $\phi$ is traceless matrix. The equation of motion of $B^{\mu\nu }_i$ is
\[
 F^i_{\mu\nu }  +\phi{^i}_j B^j_{\mu\nu } +\frac{\delta L_{matter}}{\delta B^{\mu\nu }_i}=0.
\]
The two forms $F^i_{\mu\nu }$ and $B^i_{\mu\nu }$ takes values in Lie algebra $\mathfrak{so}(3, \mathbb{C})$, so there are always matrices $\psi$ and $\psi'$ of scalar functions, so writing
\begin{equation}\label{eq:z11}
F^i=\psi{^i}_j B^j+\xi{^i}_j \bar B^j. 
\end{equation}
The equations of motion asserts that the matrices $\psi$ and $\xi$ are symmetric. 
\\

The equation of motion of the spin connection $A^i_{\mu }$ is
\begin{equation}\label{eq:z130}
-D_\mu B ^{\mu\nu i } + J^{\nu i}=0.
\end{equation}
Now we try to solve the equation (\ref{eq:z130}) using $B ^{\mu\nu i }= K{^i}_j \Sigma^{\mu\nu i }$ with $D_\mu \Sigma^{\mu\nu i }=0$. Therefore
\begin{equation}\label{eq:z16}
-(D_\mu  K{^i}_j)\Sigma^{\mu\nu i } + J^{\nu i}=0.
\end{equation}
Using $B ^{ i }= K{^i}_j \Sigma^{ i }$ in (\ref{eq:z11}), one finds that the matrix $ K$ must also be symmetric in order to satisfy the general relativity constraints using $\Sigma^{ i }$ and $F^i=(\psi K) {^i}_j \Sigma^j$ in the vacuum($\psi'{^i}_j=0$), where we let $\Sigma^i$ be a solution. By $SO(3, \mathbb{C})$ transformation using an invertable matrix $U{^i}_j$, we can get $ K{^i}_j= K^i \delta^i_j$, for some scalar functions $ K^i$. Latter we find that $ K^i$ satisfies $\partial^2  K^i=0$.
\\

The equation (\ref{eq:z16}) means that the spin current becomes a source for $D_\mu  K{^i}_j$ instead of being source for $D_\mu B ^{\mu\nu i }$. The new basis $\Sigma^{\mu\nu i }$ has to be given using gravitational field $e^I_\mu$ with the inversion $e_I^\mu$, so writing $\Sigma^{\mu\nu i }=P^{iIJ} e_I^\mu e_J^\nu$, the field $e^I_\mu$ satisfies $D_\nu e^I_\mu=0$. The matrix $ K{^i}_j$ has to be symmetric as mentioned before.
\\

Using $\Sigma^{\mu\nu i }=P^{iIJ} e_I^\mu e_J^\nu$ and $J^{\nu i}= P_{IJ}^i J^{\nu IJ}$ in (\ref{eq:z16}), we obtain
\[
\left( {D_\mu   K {^i} _j } \right)P_{IJ}^j e^{I\mu } e^{J\nu }  = P_{IJ}^i J^{\nu IJ} ,
\]
and using $J^{\nu IJ}=e^{I\mu } e^{\rho J} J_{\mu \rho }^{\nu }$ which is possible since $e_{I\mu } e_{J\rho } J^{\nu IJ}=J_{\mu \rho }^{\nu }$ exists, we find
\[
\left( {D_\mu   K {^i} _j } \right)P_{IJ}^j e^{I\mu } e^{J\nu }  = P_{IJ}^i e^{I\mu } e^{\rho J} J_{\mu \rho }^{\nu } ,
\]
so
\[
\left( {D_\mu   K {^i} _j }+v_\mu^i{ _j } \right)P_{IJ}^j e^{J\nu }  = P_{IJ}^i  e^{\rho J} J_{\mu \rho }^{\nu } ,  \quad \text{ for } v_\mu^i{ _j} P_{IJ}^j e^{I\mu }e^{J\nu }=0.
\]
Since $K {^i} _j$ is real symmetric matrix, we let $v_\mu^i{ _j } $ be also real and symmetric in $i, j$. The equation $v_\mu^i{ _j} P_{IJ}^j e^{I\mu }e^{J\nu }=0$ implies $v_\mu^i{ _j} P_{IJ}^j e^{I\mu }=0$. We can make $SO(3) \times SO(3, 1)$ transformation on $v_\mu^i{ _j} P_{IJ}^j e^{I\mu }=0$ to get $v_\mu^i{ _j}=v_\mu^i \delta^i _j$. And according to the calculations below, we find that $v_\mu^i \delta^i_j P_{IJ}^j e^{I\mu }=0$ implies $v_\mu^i =0$. So we drop $v_\mu^i{ _j }$ from now. 
\\

From the conservation $D_\nu  J^{\nu IJ}  =D_\nu  (e^{I\mu } e^{\rho J} J_{\mu \rho }^{\nu })  =e^{I\mu } e^{\rho J} D_\nu   J_{\mu \rho }^{\nu }  = 0$(using $D_\nu e^I_\mu=0$), one obtains
\[
P_{IJ}^j e^{J\nu } D_\nu  \left( {D_\mu   K {^i} _j } \right) = P_{IJ}^i e^{\rho J}  D_\nu  J_{\mu \rho }^{\nu }  = 0,
\]
so
\[
P_{IJ}^j e^{J\nu } \left( {D_\nu  D_\mu   K {^i} _j } \right) = 0, \quad \text{ for each } I=0,1,2,3 \text{ and } i=1,2,3.
\]
The matrix $ K ^i{ _j} $ is symmetric, so we can make $SO(3) \times SO(3, 1)$ transformation using a matrix $U{^i} _j$ to get $ K ^i {_j}  =  K ^i \delta {^i} _j $, so
\[
 D_\mu   K {^i} _j  = \left( {\partial _\mu   K ^i } \right)\delta {^i} _j  +  K ^i D_\mu  \delta {^i} _j .
\]
But $D_\mu  \delta {^i} _j  =0$, thus
\[
D_\mu   K {^i} _j  = \left( {\partial _\mu   K ^i } \right)\delta {^i} _j ,
\]
Therefore
\[
P_{IJ}^i e^{J\nu } \left( {\partial _\nu  \partial _\mu   K ^i } \right) = 0 \quad \text{ for each } I=0,1,2,3 \text{ and } i=1,2,3.
\]
For $I=0$, we have $P_{0j}^i=-i\delta {^i} _j/2$, it implies $e^{i\nu }  {\partial _\nu  \partial _\mu   K ^i }  = 0$ for each $i=1,2,3$. 
And for $I=j$, we have $P_{jk}^i=\epsilon{^i}_ {jk}/2$, it implies $e^{i\nu }  {\partial _\nu  \partial _\mu   K ^j } = e^{j\nu } {\partial _\nu  \partial _\mu   K ^i } $ for each $i,j=1,2,3$. The first equation $e^{i\nu } {\partial _\nu  \partial _\mu   K ^i } = 0$ implies the proportionality $  {\partial _\nu  \partial _\mu   K ^i } \sim \epsilon{^i}_ {jk} e^j_\nu $(or $\sim \epsilon_{\nu \rho \sigma \gamma} e^{\rho i}$), using it in the second equation leads to 
\[
e^{i \nu }  {\partial _\nu  \partial _\mu   K ^j } \sim \epsilon{^j}_ {\ell k} e^{i\nu } e^\ell_\nu= \epsilon{^j}_ {\ell k} \delta^{\ell i} = -\epsilon^{ij}{_k}  
\]
which is anti-symmetric in $i , j$, this is contradict with the symmetric in the second equation $e^{i\nu }  {\partial _\nu  \partial _\mu   K ^j } = e^{j\nu } {\partial _\nu  \partial _\mu   K ^i } $. Therefore we choose $e^{I\nu } \left( {\partial _\nu  \partial _\mu   K ^i } \right) = 0$, so
\begin{equation}\label{eq:z23}
e^{\mu }_I e^{I\nu } \left( {\partial _\nu  \partial _\mu   K ^i } \right)  =g^{\mu \nu } \partial _\nu  \partial _\mu   K ^i  = \partial ^2  K ^i  = 0,
\end{equation}
where we used the metric $g_{\mu \nu }=e^J_\mu  e_{J\nu }$, it satisfies $\nabla _\rho g_{\mu \nu }=(D_\rho e^J_\mu)  e_{J\nu }+ e^J_\mu  (D_\rho e_{J\nu })=0$. 
\\

We note that we used a metric $g_{\mu \nu}$ defined by $e_{J\mu}  e^J_\nu $, but the equation $\partial ^2  K ^i  =0$ is invariant under any coordinates transformation $x\to x'$, ${\psi'} ^i(x')=\psi ^i(x)$, so the equation $\partial ^2  K ^i  =0$ is valid for any other metric ${g'}_{\mu \nu}$(at least locally). Also we let the affine connection $\nabla _\mu$ be flat connection, therefore we can make coordinates transformation to obtain euclidean metric on $M$, and so $\nabla _\mu=\partial_\mu$. Another treatment of $e^{I\nu } {\partial _\nu  \partial _\mu   K ^i }= 0$ is multiplying it with $e_{I\rho }$ and sum over $I$ to get $  {\partial _\rho  \partial _\mu   K ^i }  = 0$ which can be solved by letting $K ^i$ depend only on one coordinate(like $r$ in spherical coordinates), we obtain $\eta^{\mu \nu} {\partial _\mu  \partial _\nu   K ^i }  = 0$, for Lorentz metric $\eta$. By that the functions $K ^i$ do not depend on the metric $e_{J\mu}  e^J_\nu $ on $M$, but they have to depend only on one coordinate on $M$.
\\

To get a solution, we first solve $ \partial ^2  K ^i  = 0$(by letting $K ^i$ depend only on one coordinate), then using it in $-(\partial_\mu  K^i)\Sigma^{\mu\nu i } + J^{\nu i}=0$, the equation (\ref{eq:z16}), to get the solution of $\Sigma^i_{\mu \nu }$ in terms of given spin current $ J^{\nu i}$, and so getting $B^i_{\mu \nu }$, where the spin current $J^{i \mu}$ is assumed to be given as a function on $M$. Using the solution of $\Sigma^i_{\mu \nu }$, we obtain the connection $A \in \Omega^1\left(M; \mathfrak{so}(3, \mathbb{C})\right)$ using $D \Sigma^i=0$. So obtaining the curvature $F^i(A)=dA^i +\varepsilon {^i}_{jk} A^j  \wedge A ^k$.
\\

In $3+1$ decomposition of the space-time manifold $M=\Sigma \times \mathbb{R}$, let $\Sigma_t$ be the space-like slice of constant time $t$ with the coordinates $(x^a)_{a=1,2,3}$ (and $0$ is time index). The equation $ D_\mu \Sigma^{ \mu \nu i}=0$ decomposes to two equations, 
\begin{equation}\label{eq:z29}
D_a E^{a  i}=0  \quad\text{ and  }\quad  \epsilon^{abc} D_b  B_c^{ i}=0,  \quad\text{ for   } \quad \epsilon^{abc}=\epsilon^{0abc},
\end{equation}
in which we introduce the vector field $E^i$ and the 1-form $B^i$, 
\begin{equation}\label{eq:z24}
E^{a  i}= \Sigma^{ 0 a i}, \quad\text{ and  } \quad \Sigma^{ ab i}=\epsilon^{abc} B_c^{ i},
\end{equation}
on the space-like slice $\Sigma_t$(the field $E^{a  i}$ is conjugate to the connection $A^i_a$). 
\\

By that the equation $-(\partial_\mu  K^i)\Sigma^{\mu\nu i } + J^{\nu i}=0$ yields (with $\partial_0  K^i=0$)
\begin{equation}\label{eq:z22}
\begin{split}
& J^{a i}=(\partial_b  K^i)\Sigma^{b a i }=(\partial_b  K^i)\epsilon^{bac} B_c^{ i}=\epsilon^{bac}(\partial_b  K^i) B_c^{ i},\\
and &\\
& J^{0 i}=(\partial_a  K^i)\Sigma^{a 0 i }=-(\partial_a  K^i)\Sigma^{0a i }=-(\partial_a  K^i) E^{a  i},
\end{split}
\end{equation}
where $J^{a i}$ are vector currents and $J^{0 i}=Q^i$ are charges. 
\\

In the static case $J^{a i}=0$, we can solve the first equation in terms of $\partial_a  K ^i$ by writing 
\begin{equation}\label{eq:z26}
B_a^{ i}=f(\partial_a  K^i) b^i, \quad\text{ for   } b^i \in \Gamma\left(M; \mathfrak{so}(3, \mathbb{C})\right),
\end{equation}
and $f$ is scalar function on $M$. But when $J^{a i}\ne 0$, we let $B_a^{ i}=f(\partial_a  G^i) b^i$ for some scalar functions $G^i \ne K^i$. The equation $\epsilon^{abc} D_b  B_c^{ i}=0$ implies $D b^i=0$ and $df=0$.
\\

Regarding the second equation of (\ref{eq:z22}), when $J^{0 i}=0$(in the vacuum), we can solve it in terms of $\partial_a  K ^i$ by writing 
\begin{equation}\label{eq:z27}
E^{a  i}=g\epsilon^{abc}(\partial_b  K^i) v^i_c,
\end{equation}
for some 1-form $v \in \Gamma\left(M; T^*\Sigma \times \mathfrak{so}(3, \mathbb{C}\right)$, and $g$ is scalar function on $M$. The equation $D_a E^{a  i}=0$ implies $D v=0$ and $dg=0$. 
\\

And when $J^{0 i}\ne 0$(existence of matter), we let $E^{a  i}=g^{ab} (\partial_b  K^i) u^i$, for some vector field $u \in \Gamma\left(M; \mathfrak{so}(3, \mathbb{C})\right)$. By that 
\begin{equation}\label{eq:z17}
\begin{split}
Q^i= J^{0 i}=-g^{ab}(\partial_a  K^i) (\partial_b  K^i) u^i.
\end{split}
\end{equation}
Thus there are two solutions of $E^{a  i}$(for vacuum and for matter), the total solution is $E^{a  i}=g^{ab} (\partial_b  K^i) u^i + g\epsilon^{abc}(\partial_b  K^i) v^i_c$. The equation $D_a E^{a  i}=0$ implies $D u=0$ since $\partial^2  K^i=0$ as showed before.
\\

Using the self-duality \cite{Felix, Bennett}
\begin{equation}\label{eq:z18}
\frac{1}{{2!}}e^{-1} \varepsilon^{\mu \nu \rho \sigma } \Sigma _{\rho \sigma }^i    =\left( {*\Sigma ^i } \right)^{\mu \nu }  = \left( { - i\Sigma ^i } \right)^{\mu \nu }  =  - i\Sigma ^{\mu \nu i }, \quad e=\det(e_\mu^I),
\end{equation}
where we used the Hodge duality theory between the forms and the tensor fields, here $\Sigma^i_{\mu\nu}$ is 2-form and $\Sigma^{i\mu\nu}$ is (0,2) tensor field. Regarding the equations (\ref{eq:z29}), (\ref{eq:z26}),  (\ref{eq:z27}) and  (\ref{eq:z18}), one can take the solutions
\begin{equation}\label{eq:z19}
\begin{split}
\Sigma_{0a}^i=-\Sigma_{a0}^i=(\partial_a  K^i) b^i, \quad\text{ and   }\quad \Sigma_{ab}^i=-\Sigma_{ba}^i=(\partial_{[a}  K^i) v^i_{b]}.
\end{split}
\end{equation}

In order to write $\Sigma^{IJ}$ as $e^I \wedge e^J$ in the solutions (\ref{eq:z19}), as required to get gravity theory, we have to let $v^i_a=v_a b^i$, for some 1-form $v \in \Gamma\left(M; T^*\Sigma\right)$. Therefore the solutions (\ref{eq:z19}) become
\begin{equation}\label{eq:z20}
\begin{split}
&\Sigma_{0a}^i=-\Sigma_{a0}^i=(\partial_a  K^i) b^i, \quad\text{ and   }\quad \Sigma_{ab}^i=-\Sigma_{ba}^i=(\partial_{[a}  K^i) v_{b]}b^i,\\
\text{ for   }&\\
& b^i \in \Gamma\left(M; \mathfrak{so}(3, \mathbb{C})\right),\quad Db^i =0, \text{ and } dv=0.
\end{split}
\end{equation}

And because of $v_a E^{a  i}=0$, we let $v^a=g^{ab} v_b$ be Killing vector satisfying $v^a \partial_a  K^i=0$, so $v^a B^i_a=0$. By that and according to self-dual projection, there is vector fields $b^I$ and $ K^I$ satisfying $ K^i b^i=P^i_{IJ} b^I ( K^J b^J)$, therefore $\Sigma^{IJ}_{ab}=\left(b^I v_{[a} \right)D_{b]}\left(  K^J b^J\right) $, then we can write $e^I_a=v_{a}b^I$ and $e^J_b=\left(\partial_{b}  K^J\right)b^J$. And from $\Sigma_{0a}^{IJ}=u^I \left(\partial_a  K^J \right)u^J$, we get $e^J_a=\left(\partial_{a}  K^J\right)u^J$ and $e^I_0=v_0 u^I$, for $v_0=1$. A more general case is to find three vector fields $b_1^I$, $b_2^I$ and $ K^I$ satisfying $ K^i b^i=P^i_{IJ} b_1^I ( K^J b_2^J)$, therefore $\Sigma^{IJ}_{ab}=\left(b_1^I v_{[a} \right)D_{b]}\left(  K^J b_2^J\right) $, then we can write $e^I_a=v_{a}b_1^I$ and $e^J_b=\left(\partial_{b}  K^J\right)b_2^J$. And from $\Sigma_{0a}^{IJ}=u_1^I \left(\partial_a  K^J \right)u_2^J$, we get $e^J_a=\left(\partial_{a}  K^J\right)u_2^J$ and $e^I_0=v_0 u_1^I$, for $v_0=1$. By that the solutions (\ref{eq:z20}) can be written as $\Sigma^i=P^i_{IJ} \Sigma^{IJ}$ for $\Sigma^{IJ}=e^I \wedge e^J$. But we have to note that the solution (\ref{eq:z20}) is a general solution and we have to find a special solution, like to choose $b^i=u^i$ and let it be constant field, as we will do in the following study.
\\

If the charges $J^{0 i} \ne 0$ are given as a functions on $M$, and in order to get a solution using them, we use the solutions of $ K^i$(obtained from $\partial^2  K^i=0$) in the equation (\ref{eq:z17}), so obtaining the field $u^i=b^i$. And obtaining the vector $v$ from $v^a \partial_a  K^i=0$ so obtaining the 1-form by $v_a=g_{ab} v^b$. We obtain $\Sigma^{i}_{0a}$ and $\Sigma^{i}_{ab}$ using the solutions (\ref{eq:z20}), so getting the connection $A^i$ from $D \Sigma^i  =0$. We note that $v^a E^{i}_a=0$, $v_a B^{ai}=0$ and $v^a \partial_a  K^i=0$ depend on the symmetry of the system, for example, spherical symmetry, cylindrical symmetry, and so on. 
\\

Using the solutions (\ref{eq:z20}), we get the connection $A \in \Omega^1\left(M; \mathfrak{so}(3, \mathbb{C})\right)$ using $D\Sigma^{i}=0$, where $D=d+A$ is exterior covariant derivative, the good thing with it is that it does not include the affine connection $\nabla$ on $TM$. The equation $D\Sigma^{i}=0$ yields
\[
\varepsilon ^{\mu \nu \rho \sigma } \partial _\nu  \Sigma _{\rho \sigma }^i  + \varepsilon ^{\mu \nu \rho \sigma } \varepsilon {^i} _{jk} A_\nu ^j \Sigma _{\rho \sigma }^k  = 0.
\]
The solution of the spin connection $A^i$ is
\begin{equation}\label{eq:z30}
A_\mu ^i  =  \varepsilon _{\mu\nu \rho \sigma  } V^\nu  \Sigma ^{\rho\sigma  i} ,
\end{equation}
with
\[
V_\gamma   = \frac{{ - 1}}{4}\varepsilon ^{\mu \nu \rho \sigma } \Sigma _{i\gamma \mu } \partial _\nu  \Sigma _{\rho \sigma }^i ,\quad for\quad \varepsilon ^{ 0123 }=-1,\quad \varepsilon_{ 0123 }=1 .
\]
The used metric is $g_{\mu \nu}=e^I_\mu  e^J_\nu \eta_{IJ}$ and $\Sigma ^{\mu \nu}_i$ is inversion of $ \Sigma_{\mu \nu }^i$(see appendix A for more details). By using
\[
 - i\Sigma ^{\mu \nu i }=\frac{1}{{2!}}e^{-1} \varepsilon^{\mu \nu \rho \sigma } \Sigma _{\rho \sigma }^i , \quad e=\det(e_\mu^I),
\]
we get the inversion tensor $\Sigma ^{\mu \nu i }$, and $e_\mu^I$ is obtained from $\Sigma^{i}_{\mu\nu}=P^i_{IJ} e^I_\mu e^J_\nu$. We obtain the curvature $F^i(A)=dA^i +\varepsilon {^i}_{jk} A^j  \wedge A ^k$, then we satisfy the equation $\Sigma_i^{\mu\nu} F^i_{\mu\nu}(A)=0$ in the vacuum. Since $A_\mu ^i \sim b^i$, the equation $D b^i=0$ implies that $b^i$ is constant field.
\\

We try to give an example for obtaining that solutions. In spherical symmetry, where the matter located at a point. We use the spherically coordinates $(r, \theta, \varphi)$ on the space-like slice $\Sigma_t=\Sigma=\mathbb{R}^3$. We let the solutions depend only on the radius $r$, and start with solution of $\partial^2  K^i=0$(the equation (\ref{eq:z23})),
\begin{equation}\label{eq:1000}
\nabla^2  K^i= \frac{1}{{r^2 }}\frac{\partial }{{\partial r}}\left( {r^2 \frac{\partial }{{\partial r}} K ^i } \right) = 0 \Rightarrow  K ^i  =  \frac{{c^i }}{r},
\end{equation}
for some constants $c^i\in \mathbb{R}$. By that the functions $K ^i$ depend only on one coordinate, $r$, as discussed below of equation ((\ref{eq:z23})). Therefore 
\[
d  K ^i = c^i\left( {d r\partial _r  + d \theta  \partial _\theta   + d \varphi \partial _\varphi  } \right)\frac{{1}}{r} =  - \frac{{ c^i }}{{r^2 }}d r,
\]
thus the 1-form $v$ ($dv=0$, $g^{ab} v_a \partial_b  K ^i=0$) is
\[
v=a_1  d\theta    +a_2 d \varphi , \text{  } a_1, a_2 \in \mathbb{R},
\]
where in the spherical symmetry we let $a_1$ and $a_2$ do not depend on the coordinates $\theta$ and $\varphi$. The values of the constants $a_1$ and $a_2$ are not significant since $a^i=g^{ij} a_j$ is Killing vector, thus we set $a_1=a_2=1$. The used metric $g_{ab}$ here is the standard metric in the spherical coordinates, because $a^i$ is Killing vector of $K^i$. 
\\

We get the solutions of the 1-form $B^i$ and the vector field $E^i$ using (\ref{eq:z20}),
\[
B^i  =(d  K ^i  ) b^i =  - \frac{{b^i }}{{r^2 }}dr,  
\] 
and
\begin{equation*}
\begin{split}
E^i  = \frac{1}{2 }\varepsilon^{abc}\Sigma_{bc}^i \partial_a= \frac{1}{2 } \varepsilon^{abc} v_b  \left( \partial_c K^i \right) b^i\partial_a& = -\varepsilon ^{\varphi \theta r} \frac{{ b^i }}{{2r^2 }}\partial_\varphi  - \varepsilon ^{\theta \varphi r} \frac{{ b^i }}{{2r^2 }}\partial_\theta \\
&= \varepsilon ^{\varphi r\theta } \frac{{ b^i }}{{2r^2 }}\partial_\varphi  + \varepsilon ^{\theta r \varphi } \frac{{ b^i }}{{2r^2 }}\partial_\theta. 
\end{split} 
\end{equation*}
By that we get 
\begin{equation}\label{eq:z24}
\Sigma_{0r}^i =  - \frac{{b^i }}{{r^2 }}, \quad \Sigma_{r\theta}^i=\frac{{ b^i }}{{2r^2 }},  \quad \Sigma_{r\varphi}^i=\frac{{ b^i }}{{2r^2 }}.
\end{equation}
According to self-dual map, there are at least two constant fields $b_1^I$ and $b_2^I$ satisfying $b^i=P^i_{IJ} b_1^I b_2^J$. So from $\Sigma_{\mu \nu}^i=P^i_{IJ} e_\mu^I e_\nu^J$, we get the gravitational fields 
\begin{equation}\label{eq:z31}
e_0^I=- \frac{b_1^I }{r}, \quad e_r^I= \frac{b_2^I }{r} , \quad e_\theta^I=- \frac{b_1^I }{2r} ,  \quad e_\varphi^I=- \frac{b_1^I }{2r}.
\end{equation}

We let that matter be located at the origin $(0,0,0) \in \mathbb{R}^3$, therefore the charge (\ref{eq:z17}) is given by $Q^i(x)=Q^i_0 \delta^3(x)$, so $\int\limits_{\mathbb{R}^3} {Q^i_0 \delta ^3 (x)}  = Q^i_0=constant $(conservation of the charges). Using the equation (\ref{eq:z17})
\begin{equation*}
\begin{split}
Q^i= J^{0 i}=-g^{ab}(\partial_a  K^i) (\partial_b  K^i) u^i.
\end{split}
\end{equation*}
Using $K^i=c^i/r$, one finds
\begin{equation*}
Q^i(r)=- \frac{c^i c^i}{r^4}  u^i .
\end{equation*}
And in order to get $Q^i(x)=Q^i_0 \delta^3(x)$, we let 
\[
Q^i(x)=\frac{1}{\pi^2\sqrt{2}}\frac{\epsilon}{r^4+\epsilon^4} Q^i_0,\quad for \quad \epsilon\to 0^+.
\]
Therefore 
\[
-c^i c^i u^i =\frac{1}{\pi^2\sqrt{2}}\epsilon Q^i_0.
\]
So we choose $c^i=1$ and $ u^i =-\epsilon Q^i_0/(\pi^2\sqrt{2})$, and as we saw before the total solution of $E^{ai}$ (for vacuum and for matter) is $E^{a  i}=g^{ab} (\partial_b  K^i) u^i + g\epsilon^{abc}(\partial_b  K^i) v^i_c$, discussion below of (\ref{eq:z17}). Therefore in the limit $\epsilon\to 0^+$, the remaining solution is $E^{ai}=\epsilon^{abc}(\partial_b  K^i) v_c b^i$.
\\

There is multiplicity in the gravitational fields (\ref{eq:z31}), therefore we let the non-zero fields be
\begin{equation*}
e_0^0=- \frac{b_1^0 }{r}, \quad \quad e_r^1= \frac{b_2^1 }{r} , \quad \quad e_\theta^2=- \frac{b_1^2 }{2r} , \quad \quad e_\varphi^3=- \frac{b_1^3 }{2r}.
\end{equation*}
We obtain the connection $A^i_\mu$ using (\ref{eq:z30}), so getting the curvature $F^i(A)=dA^i +\varepsilon {^i}_{jk} A^j  \wedge A ^k$. The constants $b_1^I$ and $b_2^I$ have to be fixed in order to satisfy $\Sigma_i^{\mu\nu} F^i_{\mu\nu}(A)=0$ in the vacuum.
\\
\\

\section{Modifying by adding Yang-Mills Lagrangian}
In this section we get $D\Sigma^i=0$($De^I=0$) by adding a new term to the Lagrangian in self-dual formalism. We add Yang-Mills Lagrangian of the connection $A^{i}$ to the Lagrangian to get
\begin{equation}\label{eq:Lag4}
\begin{split}
L&=L_{gravity}(\Sigma, A)+L_{Y-M}(A)+ L_{matter}\\
&=\Sigma^{\mu\nu }_i F^i_{\mu\nu }  +\phi_{ij} \Sigma^{\mu\nu i}\Sigma^j_{\mu\nu } +\frac{k}{4} F^{\mu\nu }_i F^i_{\mu\nu }+ L_{matter},
\end{split}
\end{equation}
the equation of motion of $A^i_\nu$ is ($k$ is constant)
\[
- P_{iIJ} D_\mu \Sigma^{\mu\nu i} - k D_\mu F ^{\mu\nu i} + J^{\nu i}=0.
\]
Thus we can choose $D_\mu\Sigma ^{\mu\nu i}=0$, by that we get
\begin{equation}\label{eq:40}
- k D_\mu F ^{\mu\nu i} + J^{\nu i}=0, \quad \text{for}\quad  D=\nabla+A.
\end{equation}
This is same equation of motion in Yang-Mills theory of gauge fields. This relates to the fact that we can regard the local Lorentz symmetry (internal symmetry) as a gauge symmetry with spin connection $\omega^{IJ}$ (or $A^i$) as a gauge field. The Lorentz current in \ref{eq:40} is conserved since
\[
D_\nu  D_\mu F ^{\mu\nu i}= \frac{1}{2}[D_\nu,  D_\mu ] F ^{\mu\nu i}= -\frac{1}{2} \varepsilon {^i}_{jk} F _{\mu\nu }^j F ^{\mu\nu k}=0 \to D_\nu J^{\nu i}=0.
\]

We can get same current by using Riemann curvature tensor $R^\mu{_{ \nu \rho \sigma}}$ given in terms of the Levi-Civita connection $\nabla$. The Bianchi identity is
\[
\nabla_\gamma R^\mu{_{ \nu \rho \sigma}}+\nabla_\rho R^\mu{_{ \nu  \sigma\gamma}}+\nabla_\sigma R^\mu{_{ \nu  \gamma\rho}}=0.
\]
Multiplying it by $\delta^\rho_\mu$, yields to
\[
\nabla_\gamma R^\mu{_{ \nu \mu \sigma}}+\nabla_\mu R^\mu{_{ \nu  \sigma\gamma}}+\nabla_\sigma R^\mu{_{ \nu  \gamma\mu}}=0,
\]
or
\[
\nabla_\gamma R^\mu{_{ \nu \mu \sigma}}+\nabla_\mu R^\mu{_{ \nu  \sigma\gamma}}-\nabla_\sigma R^\mu{_{ \nu  \mu\gamma}}=0,
\]
the Ricci tensor $R_{ \nu  \sigma}=R^\mu{_{ \nu \mu \sigma}}$ vanishes in the vacuum, so $\nabla_\mu R^\mu{_{ \nu  \sigma\gamma}}$ also vanishes in the vacuum, so $\nabla^\mu R_{\mu \nu  \sigma\gamma}=\nabla^\mu R_{\sigma\gamma \mu \nu  }\ne0$ associates only with matter. The curvature of the spin connection $\omega$ relates with the Riemann curvature by 
\[
R_{\mu \nu}^{IJ}(\omega(e))=e^{I\sigma} e^{J\gamma} R_{\sigma\gamma \mu \nu}(g(e)), \quad \text{with}\quad g_{\mu \nu}(e)=\eta_{IJ} e^I_\mu e^J_\nu,
\]
therefore
\begin{equation*}
\nabla^\mu R_{\sigma\gamma \mu \nu  }\ne0(=0) \to D^\mu R_{\mu \nu}^{IJ}\ne0(=0), \text{  }  \text{  } D_\mu e^I_\nu=0.
\end{equation*}
Thus $D^\mu R_{\mu \nu}^{IJ}$ associates only with matter and vanishes in the vacuum. Using the self-dual projection, $F^i_{\mu\nu}(A)=P^i_{IJ}R_{\mu \nu}^{IJ}(\omega)$, we find that
\[
 J^i_{\nu}=D^\mu F^i_{\mu\nu}(A)=P^i_{IJ}D^\mu R_{\mu \nu}^{IJ}(\omega)
\]
also vanishes in the vacuum. Therefore this current associates only with matter and the equation (\ref{eq:40}) is well defined. \\

In Plebaniski formalism, we have the formula 
\[
F^i(A)=\psi{^i}_j \Sigma^j+\bar\psi{^i}_j \bar \Sigma^j, \quad \text{with}\quad \Sigma^i=P^i_{IJ} e^I \wedge e^J,
\]
in the space $(\Sigma^i, F^i)$, where $Tr(\psi)=0$, $\bar\psi{^i}_j =0$ in the vacuum without cosmological constant \cite{Kirill}. By using this formula in Equation (\ref{eq:40}), we get (Appendix B)
\begin{equation}\label{eq:J2}
(D_\mu \psi{^i}_j ) \Sigma^{j\mu\nu}=\frac{3}{ 2}J^{i\nu}  ,\quad \text{for}\quad D_\mu\Sigma^{j\mu\nu}=0.
\end{equation}
This is same equation (\ref{eq:z16}), so its solutions are same solutions we obtained in the previous section. By that we obtain the solutions of $\Sigma^i$ and $F^i(A)$, and by inserting them in the equation of motion $\delta S/ \delta \Sigma^i=0$, we obtain the Lagrangian multiplier functions $\phi_{ij}$.

\section{Decomposition Space--Time Manifold $M$ into $ R\times \Sigma$}
The formulation of general relativity based on decomposition space--time manifold $M$ into $ R\times \Sigma$ is needed for expressing the metric of space--time as a solution of an equation for time evolution, such as in the Hamiltonian formulation. Thus the time evolution is the changing of the geometry of this surface. This decomposition preserves the continuous symmetries (gauge invariance and diffeomorphism invariance) of general relativity and its canonical quantization, so we can use it for the gauge theory of general relativity \cite{Arnowitt, Rosas, Robert, Ponomarev}.
\\

We get the energy density $H$ and we set $\frac{d}{dt} H=0$ when there is no external source. We see that $\frac{d}{dt} H=0$ is same $d \theta=0$ for 3-form $\theta$ on $\Sigma$. We define gravitational field as a one-form $e^I=e^I_\mu (x) dx^\mu$ that is related with metric $g_{\mu\nu}(x)$ on an arbitrary space--time manifold $M$ by $g_{\mu\nu}=\eta_{IJ} e^I_\mu e^J_\nu$, with spin connection $ \omega^{IJ}(x)\in \Omega^1(M, \mathfrak{so}(3, 1))$, where $\mathfrak{so}(3, 1)$ is Lie algebra of Lorentz group $SO(3, 1)$. The spin connection defines covariant derivative $D_\mu$ that acts on all fields which have Lorentz indices $(I, J, ...)$:
\[
D_\mu v^I=\partial_\mu v^I + \omega^{I}_{\mu J} v^J.
\]

We start with the general relativity Lagrangian of the form
\begin{equation}\label{eq:1}
L(e, \omega)=(16\pi G)^{-1} e_I^\mu e_J^\nu (R_{\mu\nu})^{IJ}e,
\end{equation}
where
\[
 R^{IJ}=d\omega^{IJ}+\omega{^I}_K \wedge \omega^{KJ}
\]
is the Riemannian curvature tensor and $e_I^\mu$ satisfies $e^I_\mu e_J^\mu=\delta^I_J$.

By the decomposition $M \to R\times \Sigma$, we decompose this Lagrangian into
\begin{equation}\label{eq:4}
L(e, \omega)=(16\pi G)^{-1} e_I^a e_J^b (R_{ab})^{IJ}e+(16\pi G)^{-1} e_I^0 e_J^a (R_{0a})^{IJ}e.
\end{equation}
The part
\begin{equation}\label{eq:z5}
H=(16\pi G)^{-1} e_I^a e_J^b (R_{ab})^{IJ}e
\end{equation}
takes the role of the energy density on the surface $\Sigma_t$ of constant time, it does not include the time derivative, but it includes the spatial derivatives with respect to coordinates on $\Sigma_t$. It is invariant under the $SO(3,1)$ transformation, thus we let $\frac{d}{dt} H=0$ when there are no external sources. The second term
\[
e_I^0 e_J^a (R_{0a})^{IJ}e=e e_I^0 e_J^a \left(\partial_0 \omega_a^{IJ}-\partial_a \omega_0^{IJ}+\omega{^I}_{0K} \omega^{KJ}_a \right),
\]
has the term $e e_I^0 e_J^a \partial_0 \omega_a^{IJ}$ with represents the variables of the phase space $\left(e e_I^0 e_J^a , \omega_a^{IJ} \right)$ on the surface $\Sigma_t$, where we let $e e_I^0 e_J^a=E^a_{IJ}$ be conjugate momentum to $\omega_a^{IJ}$. The final term is
\[
e e_I^0 e_J^a \left(-\partial_a \omega_0^{IJ}+\omega{^I}_{0K} \omega^{KJ}_a - \omega{^I}_{aK} \omega^{KJ}_0 \right).
\]
Integrating by parts, this term becomes
\begin{equation*}
\begin{split}
 & \omega_0^{IJ} \partial_a \left(e e_I^0 e_J^a \right)+\left(\omega^I_{0K} \omega^{KJ}_a - \omega^I_{aK} \omega^{KJ}_0 \right)e e_I^0 e_J^a\\
& = \omega_0^{IJ} \partial_a \left(e e_I^0 e_J^a \right)+\left(\omega^I_{0K} \omega^{KJ}_a  -\omega^{JK}_0 \omega_{Ka}{^I}  \right)e e_I^0 e_J^a  \\
& = \omega_0^{IJ} \partial_a \left(e e_I^0 e_J^a \right)+\omega^{IK}_{0} \omega_{Ka}{^J}\left( e e_I^0 e_J^a  \right)+\omega^{KJ}_0 \omega_{Ka}{^I}  \left(e e_I^0 e_J^a \right) \\
& = \omega_0^{IJ} \partial_a \left(e e_I^0 e_J^a \right)+\omega^{IJ}_{0} \omega_{Ja}{^K} \left(e e_I^0 e_K^a \right) +\omega^{IJ}_0 \omega_{Ia}{^K}  \left(e e_K^0 e_J^a \right) \\
&=\omega_0^{IJ}D_a \left(e e_I^0 e_J^a \right).
\end{split}
\end{equation*}

The variable $\omega_0^{IJ}$ has no time derivative, so it is Lagrangian multiplier, it gives the constraint $D_a \left(e e_I^0 e_J^a \right)=D_a E^a_{IJ}=0$.
\\

In self-dual formalism of GR, we use the self-duality \cite{Felix, Bennett}
\begin{equation}\label{eq:a9}
\frac{1}{{2!}}e^{-1} \varepsilon^{\mu \nu \rho \sigma } \Sigma _{\rho \sigma }^i    =\left( {*\Sigma ^i } \right)^{\mu \nu }  = \left( { - i\Sigma ^i } \right)^{\mu \nu }  =  - i\Sigma ^{\mu \nu i }, \quad e=\det(e_\mu^I).
\end{equation}
The Lie algebra $\mathfrak{so}(3,1)$(with indices $I, J,...$) is replaced by the Lie algebra $\mathfrak{so}(3,\mathbb{C})$(with indices $i, j,...$). The field $E^a_{IJ}$ becomes complex given by \cite{Carl, Kirill}
\begin{equation}\label{eq:2}
E^{ia}=\frac{1}{2} \varepsilon^{abc}P^i_{IJ} e^I_b e^J_c , \text{ with } E^i_{ab}=\Sigma^i_{ab}=P^i_{IJ} e^I_a e^J_b
\end{equation}
where $P^i_{IJ}$ is a self-dual map given by
\begin{equation}\label{eq:selfd}
P^i_{IJ}=\frac{1}{2}\varepsilon ^i_{jk}, \text{ for } I=i, J=j, \text{ and } P^i_{0j}=-P^i_{j0}=\frac{i}{2}\delta^i_j, \text{ for } I=0, J=j \ne 0.
\end{equation}
For example $E^{1a}=\frac{1}{2} \varepsilon^{abc} ( e^2_b e^3_c+ie^0_b e^1_c)$. This map relates to the decomposition of Lie algebra of the Lorentz group $SO(1, 3)$ into two copies of Lie algebra of $SL(2, R)$ \cite{Bennett}. \\

In the equation (\ref{eq:2}), we used the formula (\ref{eq:a9}) to get
\[
\varepsilon^{0abc}P^i_{IJ} e^I_b e^J_c/2=-ie P^i_{IJ} e^{0I} e^{aJ} .
\]
Regarding the previous definition of the conjugate momentum $E^a_{IJ}=e e_I^0 e_J^a $ of $\omega_a^{IJ}$, we ignore $-i$ and just write $E^{ia}=e P^i_{IJ} e^{0I} e^{aJ}$ which becomes conjugate momentum to complex connection $A^i$.
\\

The obtained connection $A^i$ is a three-complex one-form given by
\begin{equation}\label{eq:6}
\omega^{IJ} \mapsto A^i=P{^i} _{IJ} \omega^{IJ},
\end{equation}
The curvature which associates with this connection is
\[
F^i=dA^i + \varepsilon{^i }_{jk} A^j\wedge A^k .
\]
On the surface $\Sigma_t(\sigma ^a)$, it is
\[
F^i_{ab}=\frac{1}{2} \left(\partial_b  A_b^i  - \partial_b A_a ^i + \varepsilon{^i}_{jk} A_a^j A_b^k\right) .
\]
Thus we have self-dual plus anti-self-dual map:
\begin{equation}\label{eq:21}
e^a_I e^b_J R(\omega)^{IJ}_{ab} \to \varepsilon^{abc}E_{ia} F^i_{bc}+\varepsilon^{abc}{\bar E}_{ia} {\bar F}^i_{bc},
\end{equation}
where ${\bar E}_{ia}$ and $ {\bar F}^i_{bc}$ are the Hermitian conjugate of ${ E}_{ia}$ and $ { F}^i_{bc}$. 
\\

For the Lagrangian part $H$ on $\Sigma_t$, the equation (\ref{eq:z5}), it becomes
\begin{equation} \label{eq:3}
H=\varepsilon ^{abc} E _{ai} F^i_{bc}(A)e,
\end{equation}
or
\[
Hd^3\sigma= \varepsilon ^{abc} E _{ai} F^i_{bc}(A)ed^3\sigma,
\]
where $\sigma ^a$ is coordinates on $\Sigma_t$. We can write $Hd^3\sigma$ as 3-form on $\Sigma_t$, like 
\begin{equation}\label{eq:z8}
\theta=Hd^3\sigma= E _{i} \wedge F^i(A).
\end{equation}
By that the condition $\frac{d}{dt} H=0$ becomes 
\[
({\Sigma(\sigma ^a)},d\theta)=dt \frac{d}{dt} H=0.
\]
Here $(\Sigma , V )$ is a pairing of 4-form $V\in \wedge^4 T^*_p M$ with a surface $\Sigma$, the pairing of $V$ with tangent basis in $T_p\Sigma$, defined below in equation (\ref{eq:pairing}). 
\begin{equation}\label{eq:z6}
\left( {\Sigma(\sigma ^a)} , dt\wedge \frac{d}{dt}\theta\right) =dt\frac{d}{dt}H=0.
\end{equation}
Since $\theta$ is three-form on $\Sigma(\sigma ^a)$, so $d\sigma^a \wedge \frac{\partial}{\partial \sigma^a}\theta=0$, if we add it to the last formula, we get
\begin{equation}\label{eq:z7}
\left( {\Sigma(\sigma ^a)} , dt\wedge \frac{\partial}{\partial t}\theta+d\sigma^a \wedge \frac{\partial}{\partial \sigma^a}\theta\right) = \left( {\Sigma(\sigma ^a)} , d\theta\right) =0.
\end{equation}
Under arbitrary transformation $(t, \sigma^a)\to x^\mu$, $d\sigma^a=\frac{\partial \sigma^a}{\partial x^\mu} dx^\mu$, the basis on $\Sigma$ transforms as
 \begin{equation}\label{eq:tras}
\partial_a \wedge \partial_b \wedge \partial_c   \to    \frac{\partial x^\mu}{\partial \sigma^a}\frac{\partial x^\nu}{\partial \sigma^b}\frac{\partial x^\rho}{\partial \sigma^c}\partial_\mu \wedge \partial_\nu \wedge \partial_\rho.
\end{equation}

Therefore, the components of three-form $\theta$ transforms as
\[
\theta_{abc} \to \theta_{\mu\nu\rho} =\theta_{abc}\frac{\partial \sigma^a}{\partial x^\mu} \frac{\partial \sigma^b}{\partial x^\nu} \frac{\partial \sigma^c}{\partial x^\rho} .
\]
To keep the invariance under this transformation, we let the equation (\ref{eq:z7}), $\left( \Sigma , d\theta\right)=0$ still hold. We write the components of the curvature as
\begin{equation}\label{eq:10}
F_{ab}^i=\frac{1}{2} \left(D_a  A_b^i  - D_b A_a ^i \right)=\frac{1}{2} \left(\partial_a  A_b^i  - \partial_b A_a ^i + \varepsilon{^i}_{jk} A_a^j A_b^k    \right)  ,
\end{equation}
where
\[
D_a  A_b^i=\partial_a  A_b^i + \frac{1}{2}\varepsilon{^i }_{jk} A^j_a A^k_b,
\]
which motivates introducing a notation of the covariant derivative like \cite{Rosas}
\[
D V^i=dV^i + \frac{1}{2}\varepsilon{^i }_{jk} A^j\wedge V^k ,
\]
or
\[
 D_\mu V^i_\nu=\partial_\mu V^i_\nu + \frac{1}{2}\varepsilon{^i }_{jk} A^j_\mu V^k_\nu
=\partial_\mu V^i_\nu - \frac{i}{2} A^j_\mu (T^j_A)^{ik}  V^k_\nu,
\]
so
\[
 D_\mu =\partial_\mu  - \frac{i}{2} A^j_\mu (T^j_A),
\]
where the matrix elements $(T^j_A)^{ik}=-i\varepsilon^{jik}$ are the elements of the generators $T^j_A$
in the adjoint representation of the group $SU(2)$ \cite{Howard}. The coupling constant here is $g=1$. In general we write this covariant derivative as
\begin{equation}\label{eq:7}
 D_\mu =\partial_\mu  - \frac{i}{2} g A^j_\mu (T^j_A).
\end{equation}
By that, we write the 3-form (\ref{eq:z8}) as
\begin{equation}\label{eq:8}
\theta(E , A)  =(16\pi G)^{-1/2}  E _i  \wedge DA^i .
\end{equation}
Its pairing with the surface $\Sigma_t(\sigma ^a)$ is the energy
\[
H=(\Sigma_t(\sigma ^a),\theta)=(16\pi G)^{-1/2}  \varepsilon ^{abc} {E_{ia}} F^i_{bc}(A) .
\]
Regarding the equations (\ref{eq:z6}) and (\ref{eq:z7}), we let $(\Sigma_t(\sigma ^a),d\theta)=0$, this gives a continuity equation, as we see next.
\\

Our condition $\left( {\Sigma(\sigma ^a)} , d\theta\right) =0$ makes sense here because of the decomposition $R \times\Sigma$ and fixing a coordinate system $\sigma^a$ on the hypersurface $\Sigma$, this yields to an equation of continuity on this surface. For this purpose, we take the pairing of the four-form $d\theta$ with a tangent basis on the surface $\Sigma_t(\sigma ^a)$ at an arbitrary point, we get a one-form co-vector $(\Sigma(\sigma ^a),d\theta)  $ in the direction of the normal to this surface at that point. Then we set $(\Sigma(\sigma ^a),d\theta) =0$, we obtain an equation of continuity on $\Sigma_t(\sigma ^a)$.\\

Now taking the exterior derivative of Equation (\ref{eq:8}), we obtain
\begin{equation}\label{eq:9}
(16\pi G)^{1/2} d\theta  = d\left( {E _i  \wedge DA^i } \right) = \left( {DE _i } \right) \wedge DA^i  - E _i  \wedge DDA^i ,
\end{equation}
where $E _{0i}=0$. The tri-tangent basic on $\Sigma_t(\sigma ^a)$ is $ \partial _a \wedge \partial _b \wedge \partial _c$, we rewrite it as $(1/3!)\varepsilon ^{abc} \partial _a \partial _b \partial _c$. The pairing of $d\theta$ with this basis is
\[
(16\pi G)^{1/2}\left(d\theta , \varepsilon ^{abc} \partial _a \partial _b \partial _c\right)= \left(  ( {DE _i }) \wedge DA^i, \varepsilon ^{abc} \partial _a \partial _b \partial _c\right) - \left( E _i  \wedge DDA^i , \varepsilon ^{abc} \partial _a \partial _b \partial _c\right)
\]
where $\left(\cdot , \cdot\right)$ is contraction pairing defined by
\begin{equation}\label{eq:pairing}
\left(V_{\mu\nu\rho\sigma} dx^\mu   \wedge dx^\nu   \wedge dx^\rho   \wedge dx^\sigma ,  \varepsilon ^{abc}\partial _a \partial _b \partial _c \right)= \varepsilon ^{abc}V_{\mu\nu\rho\sigma}dx^{[\mu}\delta ^\nu _c\delta ^\rho _b\delta ^{\sigma]} _a,
\end{equation}
where the bracket $[....]$ is anti-symmetrization of the indices. Although $DDA^i$ is zero when the sum is over all indices like $\varepsilon ^{abc} D_a D_b A_c^i=0$, but here we do not sum over all indices of $DDA^i$ alone, since we take common functions out of the sum.

For cotangent basis $\{dx^\mu\}$ and tangent basis $\{\partial _a\}$, this pairing can be defined simply by using inner product like \cite{Christopher}
\[
\left(dx^\mu, \partial _a\right) =\delta ^\mu _a,
\]
in which we consider $dx^a=d\sigma^a$ for $a=1, 2, 3$, so $E _{0i}=0$ regarding to our gauge.

Starting with the first term
\[
\left({( {DE _i } ) \wedge DA^i }, {\varepsilon ^{abc} \partial _a \partial _b \partial _c }\right)   =\varepsilon ^{abc}  D_\mu  E _{\nu i} D_\rho  A_\sigma ^i \left({dx^\mu   \wedge dx^\nu   \wedge dx^\rho   \wedge dx^\sigma  } ,  \partial _a \partial _b \partial _c\right)
\]
we get
\begin{align*}
&4\left({\left( {DE_i } \right) \wedge DA^i } , \varepsilon ^{abc} \partial _a \partial _b \partial _c\right) =\\
& - \varepsilon ^{abc} D_a E _{bi} D_c A_\mu ^i dx^\mu  + \varepsilon ^{abc} D_a E _{bi} D_\mu  A_c^i dx^\mu   - \varepsilon ^{abc} D_a E _{\mu i} D_b A_c^i dx^\mu  + \varepsilon ^{abc} D_\mu  E _{ai} D_b A_c^i dx^\mu  .
\end{align*}

The second term is
\[
\left( {E _i  \wedge DDA^i } ,\varepsilon ^{abc} \partial _a \partial _b \partial _c  \right)  =\varepsilon ^{abc} E _{\mu i} D_\nu  D_\rho  A_\sigma ^i \left(dx^\mu   \wedge dx^\nu   \wedge dx^\rho   \wedge dx^\sigma  , \partial _a \partial _b \partial _c \right) .
\]

Doing the same thing, we get
\begin{align*}
&4\left(E _i  \wedge DDA^i  ,\varepsilon ^{abc} \partial _a \partial _b \partial _c \right) =\\
& - \varepsilon ^{abc} E _{ai} D_b D_c A_\mu ^i dx^\mu   + \varepsilon ^{abc} E _{ai} D_b D_\mu  A_c^i dx^\mu   - \varepsilon ^{abc} E _{ai} D_\mu  D_b A_c^i dx^\mu  + \varepsilon ^{abc} E _{\mu i} D_a D_b A_c^i dx^\mu .
\end{align*}

Adding the two terms, we obtain
\begin{align*}
&4(16\pi G)^{1/2}\left(d\theta  , \varepsilon ^{abc} \partial _a \partial _b \partial _c \right)=\\
 -& \varepsilon ^{abc} D_a E _{bi} D_c A_\mu ^i dx^\mu   + \varepsilon ^{abc} D_a E _{bi} D_\mu  A_c^i dx^\mu   - \varepsilon ^{abc} D_a E _{\mu i} D_b A_c^i dx^\mu   + \varepsilon ^{abc} D_\mu  E _{ai} D_b A_c^i dx^\mu \\
 +& \varepsilon ^{abc} E _{ai} D_b D_c A_\mu ^i dx^\mu   - \varepsilon ^{abc} E _{ai} D_b D_\mu  A_c^i dx^\mu   + \varepsilon ^{abc} E _{ai} D_\mu  D_b A_c^i dx^\mu   - \varepsilon ^{abc} E _{\mu i} D_a D_b A_c^i dx^\mu  .
\end{align*}

We define the curvature by using the covariant derivative from Equation (\ref{eq:10}) as
\begin{equation}\label{eq:cur}
F_{\mu c}^i=\frac{1}{2} \left(D_\mu  A_c^i  - D_c A_\mu ^i \right)=\frac{1}{2} \left(\partial_\mu  A_c^i  - \partial_c A_\mu ^i + g\varepsilon{^i}_{jk} A_\mu^j A_c^k    \right)  ,
\end{equation}
therefore
\[
F^{\rho\nu i}=g^{\rho\mu} g^{\nu c} F_{\mu c}^i=\frac{1}{2} \left(D^\rho  A^{\nu i } - D^\nu A^{\rho i} \right);\text{ } D^\rho g^{\mu\nu }=0.
\]

Its Hodge dual on the surface $ \Sigma$ with respect to the coordinates $(\sigma^a)$ is
\[
F^{ai}=\varepsilon ^{abc} F_{bc}^i.
\]

Also we define the complex two-form field from Equation (\ref{eq:2}) as
\[
\Sigma^{bc} _{i}=E^{bc} _{i}= \varepsilon ^{bca} E _{ai}=\varepsilon ^{abc} E _{ai} .
\]

Using them in the last formula, we get
\begin{align*}
4(16\pi G)^{1/2}&\left(d\theta ,  \varepsilon ^{abc} \partial _a \partial _b \partial _c\right) =2 \varepsilon ^{abc} (D_a E _{bi}) F_{\mu c}^i dx^\mu   - 2 (D_a E _{\mu i}) F^{ai} dx^\mu       \\
&+2 (D_\mu  E _{ai}) F^{ai} dx^\mu+2E^{bc} _{i} D_b   F_{c\mu }^i dx^\mu +2E _{ai} D_\mu  F^{ai} dx^\mu  - 2 E _{\mu i} D_a F^{ai} dx^\mu  .
\end{align*}

And using
\[
\varepsilon ^{abc} (D_a E _{bi}) F_{\mu c}^i dx^\mu=-\varepsilon ^{acb} (D_a E _{bi}) F_{\mu c}^i dx^\mu=- (D_a E ^{ac}_{i}) F_{\mu c}^i dx^\mu,
\]
we obtain
\begin{align*}
 &2(16\pi G)^{1/2}\left(d\theta ,  \varepsilon ^{abc} \partial _a \partial _b \partial _c\right)  =\\
 &-  (D_a E ^{ac}_{i} )F_{\mu c}^i dx^\mu  -  (D_a E _{\mu i}) F^{ai} dx^\mu   +& (D_\mu  E _{ai}) F^{ai} dx^\mu
 -E^{bc} _{i} D_b   F_{\mu c}^i dx^\mu   \\
+&E _{ai} D_\mu  F^{ai} dx^\mu  -  E _{\mu i} D_a F^{ai} dx^\mu .
\end{align*}
With
\[
 -(D_a E ^{ac}_{i}) F_{\mu c}^i dx^\mu-E^{bc} _{i} D_b   F_{\mu c}^i dx^\mu=-D_a \left( E ^{ac}_{i} F_{\mu c}^i dx^\mu \right),
\]
it becomes
\[
2(16\pi G)^{1/2}\left(d\theta ,  \varepsilon ^{abc} \partial _a \partial _b \partial _c\right)=-D_a ( E ^{ac}_{i} F_{\mu c}^i )dx^\mu -  D_a( E _{\mu i} F^{ai} )dx^\mu + D_\mu(  E _{ai} F^{ai}) dx^\mu.
\]

As we suggested before, we let the normal of the surface $\Sigma_t(\sigma^a)$ be in direction of the time $dx^0$, so $\left(d\theta ,  \varepsilon ^{abc} \partial _a \partial _b \partial _c\right) $ is in direction of the time. Therefore we set $\mu=0$, thus we get
\[
2(16\pi G)^{1/2}\left(d\theta ,  \varepsilon ^{abc} \partial _a \partial _b \partial _c\right)=-D_a ( E ^{ac}_{i} F_{0 c}^i )dx^0 - 2 D_a( E _{0 i} F^{ai} )dx^0 + D_0( E _{ai} F^{ai}) dx^0.
\]

The vector $\left(d\theta ,  \varepsilon ^{abc} \partial _a \partial _b \partial _c\right) $ is one-form in the direction of the normal to the surface $\Sigma_t(\sigma^a)$. It is zero as we mentioned before, thus we get
\[
-D_a ( E ^{ac}_{i} F_{0 c}^i )dx^0 -  D_a( E _{0 i} F^{ai} )dx^0 + D_0( E _{ai} F^{ai}) dx^0=0,
\]
or
\[
-D_a ( E ^{ac}_{i} F_{0 c}^i ) - D_a( E _{0 i} F^{ai} ) + D_0( E _{ai} F^{ai}) =0.
\]

For $De=0$ on the surface $\Sigma$, where $e=det(e^i_a)$, We write it as 
\[
D_a ( e^2 E ^{ac}_{i} F_{c0}^i ) - D_a( E _{0 i} F^{ai} )+D_0( e^2 E _{ai} F^{ai}) =0.
\]

The term $( E _{ai} F^{ai})$ is scalar, so $D_0( E _{ai} F^{ai})=\partial_0( E _{ai} F^{ai})$, the vector $E ^{ab}_{i} F_{b0}^i$ is a usual vector field on $\Sigma$, it does not carry a Lorentz index, so $D_a ( E ^{ab}_{i} F_{b0}^i )=\partial_a ( E ^{ab}_{i} F_{b0}^i )$ and $E _{0 i}=0$, thus we get
\[
\partial_a (e^2 E ^{ab}_{i}F_{b0}^i ) + \partial_0( e^2 E _{ai} F^{ai})=0,
\]
using $F^{ai}=\varepsilon ^{abc} F_{bc}^i$, $\Sigma^{bc} _{i}=E^{bc} _{i}= \varepsilon ^{bca} E _{ai}$ and $E^{ai}=e e^{ai}$, we have  
\[
e^2 E _{ai} F^{ai}=\frac{1}{2}e^2\Sigma ^{ab}_i F_{ab}^i=\frac{1}{4}e^2 e ^{a}_i e ^{b}_j F_{ab}^{ij}=\frac{1}{4}E^{a}_i E^{b}_j  F_{ab}^{ij}=H,
\]
so we obtain
\begin{equation}\label{eq:11}
\partial_a \left(e^2 \Sigma ^{ab}_{i}F_{b0}^i \right) + \partial_0 H=2(16\pi G)^{1/2}\left(d\theta ,  \varepsilon ^{abc} \partial _a \partial _b \partial _c\right)=0.
\end{equation}
We have regarded $\frac{1}{4}E^{a}_i E^{b}_j  F_{ab}^{ij}$ as energy density. 
\\

The equation 
\[
\partial_a \left(e^2 E ^{ab}_{i}F_{b0}^i \right) + \partial_0\left( \frac{1}{4}E^{a}_i E^{b}_j  F_{ab}^{ij}\right)=0.
\]
We regard this equation as an equation of continuity with respect to a Lagrangian like $L(F^{0ai}, DA^i)$ that satisfies the action principle $\delta S(DA^i)=0$ and the invariance under continuous symmetries of general relativity. Therefore we regard $ \frac{1}{2}c\Sigma^{ab}
_i F_{ab}^i$ as energy density $T^{00}$, and $c\Sigma ^{ab}_{i} F_{b0}^i$ as momentum density $T^{0a}$, where $c$ is constant for satisfying the units.

\section{Conclusions}
We have considered the 4-manifold $M$ as a base space and consider that the Lorentz group $SO(3,1)$ act locally on Lorentz frames which are regarded as a frame bundle over a fixed base space, and consider the local Lorentz frame as an element in the tangent frame bundle over $M$. By that we have two symmetries; invariance under continuous transformations of local Lorentz frame, $SO(3,1)$ group, and invariance under diffeomorphism of the space--time $M$. The local invariance of matter Lagrangian under $SO(3,1)$ transformation produces a current, called spin current $J=\delta S_{matter}/ \delta A$, this current couples to the spin connection, so we have gauge theory, that makes the general relativity similar to Yang-Mills theory of gauge fields. We have solved the equations of motion of general relativity in self-dual formalism using only the spin currents(Lorentz currents), in static case, and without needing using the Einstein's equation. We give an example, matter located at a point, so we have spherical symmetric system. We solved the equation $\delta S/ \delta A_\nu^i=-D_\mu B ^{\mu\nu }_i + J^{\nu }_i=0$ by using $B ^{\mu\nu i }= K{^i}_j \Sigma^{\mu\nu i }$ for $D_\mu \Sigma^{\mu\nu i }=0$. 
\\

\section{Appendix A}
$\rm{I}$- We verify that the (0,2) tensor field $\Sigma ^{\mu \nu i}$ defined in(self duality)
\begin{equation}\label{eq:z21}
 - i\Sigma ^{\mu \nu i }=\frac{1}{{2!}}e^{-1} \varepsilon^{\mu \nu \rho \sigma } \Sigma _{\rho \sigma }^i , \quad e=\det(e_\mu^I) ,\quad  \varepsilon^{0123 }=-1,\quad  \varepsilon_{0123 }=1,
\end{equation}
is inversion of the 2-form $\Sigma_{\mu \nu }^i$, that is $\Sigma _{\mu \nu }^i  \Sigma ^{\mu \nu }_j=\delta^i_j$. Multiplying with $\Sigma _{\mu \nu j} $ and sum over contracted indices, we get
\[
 - i\Sigma _{\mu \nu j} \Sigma ^{\mu \nu i }=\frac{1}{{2}}e^{-1} \varepsilon^{\mu \nu \rho \sigma }\Sigma _{\mu \nu j} \Sigma _{\rho \sigma }^i .
\]
Then using $\Sigma^i_{\mu \nu }=P_{IJ}^i e^I_\mu e^J_\nu $, implies
\[
 - i\Sigma _{\mu \nu j} \Sigma ^{\mu \nu i }=\frac{1}{{2}}e^{-1} \varepsilon^{\mu \nu \rho \sigma }P_{jIJ} P^i_{KL} e^I_\mu e^J_\nu e^K_\rho e^L_\sigma.
\]
Using $\varepsilon^{\mu \nu \rho \sigma } e^I_\mu e^J_\nu e^K_\rho e^L_\sigma=e \varepsilon^{IJKL }$, to get
\[
 - i\Sigma _{\mu \nu j} \Sigma ^{\mu \nu i }=\frac{1}{{2}}e^{-1}P_{jIJ} P^i_{KL} e \varepsilon^{IJKL }=\frac{1}{{2}}P_{jIJ} P^i_{KL} \varepsilon^{IJKL }.
\]
Then we use the self-dual projection property
\[
P_{IJ}^i \varepsilon ^{IJKL}  = -2iP^{iKL}
\]
to obtain
\[
 - i\Sigma _{\mu \nu j} \Sigma ^{\mu \nu i }=\frac{1}{{2}}P_{jIJ}(-2iP^{iIJ}) =P_{jIJ}(-iP^{iIJ }).
\]
Therefore
\[
 \Sigma _{\mu \nu j} \Sigma ^{\mu \nu i } =P_{jIJ} P^{iIJ}=\delta^i_j,
\]
where we used the self-dual projection property $P_{jIJ} P^{iIJ}=\delta^i_j$. The sum is over the contracted indices.
\\

\section{Appendix B}
We calculate $D^\nu F^i_{\nu\mu}$ by using
\[
(*D*F^i)_{\mu}=D^\nu F^i_{\nu\mu},
\]
where the Hodge star operator 
\[
*F^i=* \left( \psi{^i}_j \Sigma^j+\bar\psi{^i}_j \bar \Sigma^j \right) =\psi{^i}_j (*\Sigma^j)+\bar\psi{^i}_j( *\bar \Sigma^j),
\]
is with respect to the metric $g_{\mu\nu}$. By using the properties of self-duality:
\begin{equation*}
 *\Sigma^j=-i\Sigma^j,\text{ }   *\bar\Sigma^j=i\bar\Sigma^j,
\end{equation*}
we obtain
\[
D^\nu F^i_{\nu\mu}=i\left(*(-D\psi{^i}_j\Sigma^j)+*(D\bar\psi{^i}_j \bar \Sigma^j) \right)_{\mu}.
\]
The first term becomes
\[
D(\psi{^i}_j\Sigma^j)=D(\psi{^i}_j\Sigma^j+\bar\psi{^i}_j \bar \Sigma^j)-D(\bar\psi{^i}_j \bar \Sigma^j), \text{ so } D(\psi{^i}_j\Sigma^j)=DF^i-D(\bar\psi{^i}_j \bar \Sigma^j),
\]
and by Bianchi identity $DF^i=0$, we obtain $D(\psi{^i}_j\Sigma^j)=-D(\bar\psi{^i}_j \bar \Sigma^j)$. Therefore
\begin{equation*}
D^\nu F^i_{\nu\mu}=i\left(*(-D(\psi{^i}_j  \Sigma^j)-*(D\psi{^i}_j  \Sigma^j)\right)_{\mu}=-2i(*D(\psi{^i}_j) \wedge \Sigma^j))_{\mu},
\end{equation*}
where we used $D\Sigma^j=0$. Thus we obtain
\begin{equation*}
D^\nu F^i_{\nu\mu}=-2i \frac{1}{ 3!}\epsilon_{\mu}{^{\nu \rho\sigma}} (D_\nu \psi{^i}_j)  \Sigma^j_{\rho\sigma} ,
\end{equation*}
or
\begin{equation*}
D^\nu F^i_{\nu\mu}=- \frac{2i}{ 3!}\epsilon_{\mu\nu \rho\sigma} (D^\nu \psi{^i}_j)  \Sigma^{j\rho\sigma} ,
\end{equation*}
the raising and lowering is done by using the metric $g_{\mu\nu}$. But
\[
 \frac{1}{ 2!}\epsilon_{\mu\nu \rho\sigma} \Sigma^{j\rho\sigma}=(* \Sigma^j)_{\mu\nu}=(-i\Sigma^j)_{\mu\nu},
\]
so we obtain
\begin{equation*}
D^\nu F^i_{\nu\mu}=\frac{-2i}{ 3}  (D^\nu \psi{^i}_j)  (-i\Sigma^j)_{\mu\nu} =\frac{-2}{ 3}  (D^\nu \psi{^i}_j)  \Sigma^j_{\mu\nu}=\frac{2}{ 3}  (D^\nu \psi{^i}_j)  \Sigma^j_{\nu\mu}.
\end{equation*}
From $D^\nu F^i_{\nu\mu}=J^i_\mu$, we get
\begin{equation*}
 (D^\nu \psi{^i}_j)  \Sigma^j_{\nu\mu}=\frac{3}{ 2} J^i_\mu , \text{ so } \quad  (D_\nu \psi{^i}_j)  \Sigma^{j\nu\mu}=\frac{3}{ 2} J^{i\mu}.
\end{equation*}
This is a linear equation in the basis $\{ \Sigma^i \}$.

\end{document}